# Optimization reduces knee-joint forces during walking and squatting: Validating the inverse dynamics approach for full body movements on instrumented knee prostheses

Heiko Wagner[1,2,4,a], Kim Joris Boström[1,b], Marc H.E. de Lussanet[1,2,c], Myriam L. de Graaf[1,2,d], Christian Puta[3,4,e], and Luis Mochizuki[5,f]

[1] Movement Science, University of Münster, Münster, Germany
[2] Otto Creutzfeldt Center for Cognitive and Behavioral Neuroscience, University of Münster
[3] Department of Sports Medicine and Health Promotion, Friedrich-Schiller-Universität Jena, Jena, Germany
[4] Center for Interdisciplinary Prevention of Diseases related to Professional Activities, Friedrich-Schiller-Universität, Jena, Germany
[5] School of Arts, Sciences, and Humanities, University of Sao Paulo, Brazil

[a] heiko.wagner@uni-muenster.de (Corresponding author); ORCID 0000-0002-5470-5044
[b] kim.bostroem@uni-muenster.de ; ORCID 0000-0001-5966-458X
[c] lussanet@uni-muenster.de ; ORCID 0000-0003-3754-6675
[d] mdegraaf@uni-muenster.de ; ORCID 0000-0002-9216-1121
[e] christian.puta@uni-jena.de ; ORCID 0000-0003-3936-4605
[f] mochi@usp.br ; ORCID 0000-0002-7550-2537

**Abstract**

Due to the redundancy of our motor system, movements can be performed in many ways. While multiple motor control strategies can all lead to the desired behavior, they result in different joint and muscle forces. This creates opportunities to explore this redundancy, e.g., for pain avoidance or reducing the risk of further injury. To assess the effect of different motor control optimization strategies, a direct measurement of muscle and joint forces is desirable, but problematic for medical and ethical reasons. Computational modeling might provide a solution by calculating approximations of these forces. In this study, we used a full-body computational musculoskeletal model to (1) predict forces measured in knee prostheses during walking and squatting and (2) to study the effect of different motor control strategies (i.e., minimizing joint force vs. muscle activation) on the joint load and prediction error. We found that musculoskeletal models can accurately predict knee joint forces with an RMSE of <0.5 BW in the superior direction and about 0.1 BW in the medial and anterior directions. Generally, minimization of joint forces produced the best predictions. Furthermore, minimizing muscle activation resulted in maximum knee forces of about 4 BW for walking and 2.5 BW for squatting. Minimizing joint forces resulted in maximum knee forces of 2.25 BW and 2.12 BW, i.e., a reduction of 44% and 15%, respectively. Thus, changing the muscular coordination strategy can strongly affect knee joint forces. Patients with a knee prosthesis may adapt their neuromuscular activation to reduce joint forces during locomotion.

# 1.  Introduction

Several circumstances in daily living require the minimization of forces acting on a joint. Such circumstances involve, for instance, movement or non-movement related pain, resulting from e.g., osteoarthritis, osteoporosis, inflammatory diseases of joints (such as rheumatoid arthritis), injuries of articular cartilage, etc. People experiencing pain may try to reduce the pain and/or improve movement by shifting load to the non-painful leg or changing their gait pattern. Physiotherapeutic approaches focus on pain reduction, movement enhancement and adapting the neuromuscular coordination between agonistic and antagonistic muscle activation (Skou et al., 2015), for instance, to reduce antagonistic co-activation of muscles acting on the same joint. These applied therapeutic approaches aim to develop alternative optimization strategies to balance knee joint forces and improve muscular control of the knee joint. To estimate the effect of such therapeutic interventions, it is advantageous to know, as precisely as possible, the forces in the joints and muscles for common activities, such as walking and squatting.

Although direct measurement of internal joint forces is desirable, a direct invasive measurement of joint forces is problematic due to ethical and medical issues. Only a few studies have directly measured joint forces in humans in vivo (Fregly et al., 2012; Kia et al., 2014; Trepczynski et al., 2018, 2019). Taylor et al. (2017) directly measured knee joint forces, whole body kinematics, electromyography (EMG) of some leg muscles, and ground reaction forces (GRF).

As an alternative to such invasive methods, joint forces can be approximated using musculoskeletal models. It is theoretically possible to calculate the dynamic 3D joint torques and forces when the inertial properties of the body segments, the kinematics, the external forces and the internal muscular forces are known.

Early musculoskeletal models consisted of one joint and a single muscle connected between two segments (e.g., Sherrington 1913, Sust et al. 1997, Wagner et al. 1999, Brown et al. 2000). These models were further developed to more complex 2D descriptions of human extremities (e.g., Pandy et al. 1988, Bobbert et al. 1988, van Soest et al. 1993, Guenther et al. 2003), that were able to provide more insights into motor control strategies and the importance of muscular redundancies and their contribution to stability.

Today, because of new algorithms and faster processors, it is possible to calculate 3D-full-body musculoskeletal models, with hundreds of muscles in an acceptable time (e.g., AnyBody Technology, Aalborg, Denmark; Biomechanics of Bodies (BoB) Coventry, United Kingdom; OpenSim, Stanford California, USA; Myonardo, Predimo GmbH, Münster, Germany; Walter et al. 2021; Human Body Model, Motek, the Netherlands). Even though these models have been conceived with different purposes, such as the scientific, educational and commercial goals, they are developed on a similar physical basis to describe multi-body dynamics. We developed Myonardo as a Matlab Simulink based model to calculate joint and muscle forces, and to investigate the neurophysiology and pathophysiology of motor control strategies, as well as the spinal control of human and animal locomotion.

When using these models, it is not possible to directly calculate the force contribution of each single muscle, due to the muscular redundancy of the motor system. The redundancy can be addressed by using optimization algorithms. At each time point, a muscular activity pattern is sought that not only generates the necessary joint torque, but additionally minimizes some predefined target function. Common examples of target functions are the overall muscle activation, muscle strength, joint force or the energy consumption (Dembia et al., 2021).

In the present study, we compared the influence of two optimization criteria on the predicted muscle activation and joint force, namely: minimal muscle activation and minimal joint force. Optimizing muscle activation reflects a physiologically plausible economic criterion, while optimizing resultant joint force reflects a more clinical criterion. The first criterion is more plausible in persevering activities, whereas the latter is more plausible for clinical conditions involving joint pain.

Given the approximations due to muscular redundancy and the inaccuracies of the measured kinematics and inertial properties of body segments, it is not yet clear how accurate predicted joint forces and torques are for real-life biological movements. Due to the difficulty of measuring joint loads in vivo, validation has proven problematic (Nejad et al., 2020). To validate a model, the difference between measured and calculated outputs, i.e., the error, should be small and random.

The aim of this study was to assess the effect of different optimization strategies (minimizing muscle activation vs joint forces) on the predicted knee joint forces during walking and squatting. To this end, we morphed our optimization cost function in a stepwise manner from a focus on minimizing muscle activation to minimizing resultant joint force, and we calculated the internal knee forces for each. To ensure the validity of our model, we compared our model predictions with corresponding real-life date recorded in instrumented knee prostheses (Taylor et al., 2017).

## 2.   Materials and Methods

The "Comprehensive Assessment of the Musculoskeletal System" (CAMS-knee) dataset (Taylor et al., 2017) was the experimental basis for the analyses of the present study. This dataset encompasses kinematic data, ground reaction forces and measured joint forces in the knee prostheses of six patients (1 female, 68±5 year, 88±12 kg, 173±4 cm). We used the kinematic data and the measured ground reaction forces to drive a musculoskeletal model, in order to predict the forces in the knee joint for a spectrum of optimization criteria. These predicted forces were then compared with the measured joint forces of the instrumented prostheses.

A summary of the experimental setup of the CAMS-knee study is necessary for understanding the present study. Six patients with instrumented knee implants (Zimmer, Switzerland; FIXUC) participated in the CAMS-knee study. Among others, the patients were asked to perform level walking at a self-selected speed, as well as squatting movements. The following data were measured: whole body kinematics (26 cameras, 100 fps, Vicon, UMG, UK), ground reaction forces (6 force plates, 2000Hz, Kistler Instrumentation,

Winterthur, Switzerland) and bilateral electromyography (EMG) (2000 Hz, Trigno, Delsys, USA) of the rectus femoris, vastus medialis, vastus lateralis, tibialis anterior, semitendinosus, biceps femoris long head, and the medial and lateral gastrocnemius. The medial force (positive for lateral direction), the anterior force (positive for anterior direction), and the superior force (positive for superior direction) in the prostheses were measured simultaneously at 100 Hz (Heinlein et al., 2007)

## 2.1  Computational musculoskeletal model

For the computational modeling, a full body musculoskeletal model called Myonardo (Version 3.4.4, Predimo GmbH, Münster, Germany) was used. This model is written in Matlab using the Simscape Multibody 3D simulation environment (Version 2021a, The MathWorks, Inc., Natick, Massachusetts, United States).

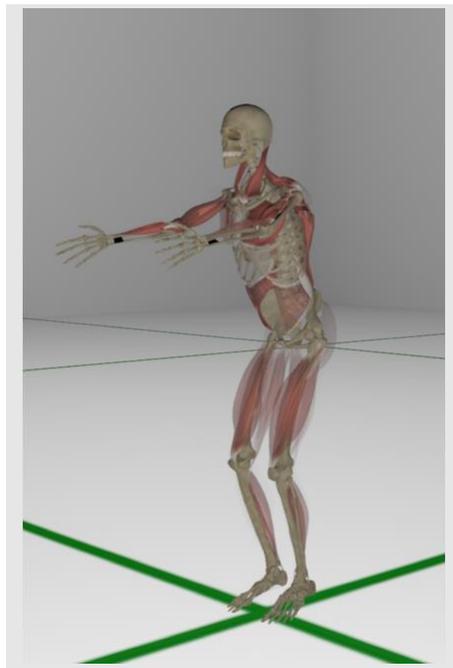

Figure 1: Visualization of the muscle activation of Myonardo (Predimo GmbH, Münster, Germany), during a squatting movement.

The whole-body model consists of 23 segments with 23 joints (66 degrees of freedom DoF), and 682 muscle-tendon units (Fig. 1). The full-body kinematics were calculated, but the subsequent analysis only focused on the lower body, thereby reducing the model to 9 segments with 9 joints (18 DoF), and 118 muscle-tendon units. The model was scaled relative to the body mass and height of the subjects (Hatze, 1980; Winter, 2009), e.g., the maximum isometric forces of each muscle scaled quadratically and the optimum muscle length linearly, relative to the body height. The relative mass and size of the segments, and the muscular attachments were selected according to Shippen and May (2012).

The musculoskeletal model was used to calculate the net joint torques and thus to predict the muscle forces acting on each joint based on an optimization algorithm. To calculate the net joint torques, the kinematics of the body segments and the external ground reaction force acting on the corresponding left and right foot segment were applied to the skeletal model. We calculated the muscle-tendon length $l_m$ as the distance between the attachment points of the muscles at the segments, the muscle-tendon velocity $v_m$ and the muscle lever arm relative to the momentary joint center, for each contributing muscle $m$. The resulting muscle force $f_m$ was calculated with a Hill-type model based on the force-length relation $f_{l,m}$ and the force-velocity relation $f_{v,m}$. The latter was formulated as (Wagner and Blickhan 2003, Thaller and Wagner 2004)

$$f_{v,m} = \begin{cases} \frac{c_m}{-v_m + b_m} - a_m & \text{if } v_m \leq 0 \\ \frac{C_m}{-v_m - B_m} + A_m & \text{if } v_m > 0 \end{cases} \tag{1}$$

where the variables $a_m, b_m, c_m$ and $A_m, B_m, C_m$ were estimated from the isometric force $f_{iso,m}$, resting length $l_{opt,m}$, eccentric gain $f_{ecc}$ and ratio of fast-twitch over slow-twitch fibers $FT$ according to

$$\begin{aligned}
a_m &= 0.25 \cdot f_{\text{iso},m} \\
b_m &= 0.25 \cdot (6 + 10 \cdot \text{FT}) \cdot l_{\text{opt},m} \\
c_m &= b \cdot (f_{\text{iso},m} + a_m) \\
A_m &= f_{\text{ecc}} \cdot f_{\text{iso},m} \\
B_m &= (A_m - f_{\text{iso},m}) \cdot b_m^2 / c_m \\
C_m &= (A_m - f_{\text{iso},m}) \cdot B_m \ .
\end{aligned} \tag{2}$$

The active and passive force-length relations were calculated as (Hatze 1981, Otten 1987)

$$f_{l,m} = \exp\left(-\left(\frac{(l_m/l_{\text{opt},m})^{k_1} - 1}{k_2}\right)^{k_3}\right), \tag{3}$$

and

$$f_{p,m} = \exp\left((l_m - l_{\text{opt},m})/k_4\right) \tag{4}$$

respectively. The resulting muscle force vector obtains

$$\vec{F}_m = act_m \cdot \vec{F}_{\text{act},m} + \vec{F}_{\text{pass},m} \ , \tag{5}$$

with the active and passive force contribution defined by

$$\vec{F}_{\text{act},m} = f_{v,m} \cdot f_{l,m} \cdot \vec{d}_m \tag{6}$$

$$\vec{F}_{\text{pass},m} = f_{p,m} \cdot \vec{d}_m , \tag{7}$$

respectively, where $\vec{d}_m$ is the normalized direction vector pointing from the origin of the muscle to its insertion, and where $act_m$ denotes the activation of muscle $m$ restricted to $act_m \in [0,1]$ for all muscles $m$ at all times. Here and in the following, $\vec{\cdot}$ denotes a spatial, 3-component vector or pseudovector. A pseudovector is a three-component object that does not always correctly transform under spatial transformations like a regular vector. For example, the angular velocity can be represented by a pseudovector $\vec{\omega}$, which does not flip its direction under spatial inversion as regular vectors do.

For the present study we used $k_1 = 0.96, k_2 = 0.35, k_3 = 2, f_{ecc} = 1.5$, and $FT = 0.5$ as global parameters (Liebetrau et al. 2013), and we took $f_{iso,m}$ and $l_{opt,m}$ for each contributing muscle $m$ from Rajagopal et al. (2016). The resulting joint torque that is generated by muscle $m$ drawing over joint $j$ is given by

$$\vec{\tau}_{j,m} = (\vec{r}_{j,m} \times \vec{F}_m) \cdot 1_{\mathcal{M}_j}(m) , \tag{8}$$

where $\vec{r}_{j,m}$ is the momentary lever arm of muscle $m$ relative to the rotation center of joint $j$, $\mathcal{M}_j$ is the set of all muscles drawing over joint $j$, and

$$1_{\mathcal{M}_j}(m) = \begin{cases} 1 & \text{if } m \in \mathcal{M}_j \\ 0 & \text{else} \end{cases} \tag{9}$$

is the indicator function of muscle $m$ being in the set $\mathcal{M}_j$. That way, we can sum over all muscles $m = 1, \dots, M$ to obtain the total torque acting on joint $j$,

$$\vec{\tau}_j = \sum_{m=1}^{M} (\vec{r}_{j,m} \times \vec{F}_m) \cdot 1_{\mathcal{M}_j}(m) . \tag{10}$$

These calculations are carried out for each point in time.

## 2.2   Optimization method

To solve the muscular redundancy, different optimization criteria can be applied to determine the activation of each contributing muscle at a given point in time. These different optimization algorithms result in different muscle activation patterns, and therefore in different muscle and joint forces. It should be stressed that the joint forces, i.e., the forces acting upon a given joint, result not just from the mechanical load, but primarily from the contraction of muscles. We compared the influence of two optimization criteria on the predicted joint forces, namely (1) minimizing the muscle activation, and (2) minimizing the resulting joint force.

Let us define the active and passive torque component of muscle $m$ acting on joint $j$ at a given point in time by

$$\vec{\tau}_{\text{act},j,m} = \left( \vec{r}_{j,m} \times \vec{F}_{\text{act},m} \right) \cdot 1_{\mathcal{M}_j}(m) \tag{11}$$

$$\vec{\tau}_{\text{pass},j,m} = \left( \vec{r}_{j,m} \times \vec{F}_{\text{pass},m} \right) \cdot 1_{\mathcal{M}_j}(m) , \tag{12}$$

respectively, so that the total torque acting on joint $j$ is given by

$$\vec{\tau}_j = \sum_{m=1}^{M} act_m \cdot \vec{\tau}_{\text{act},j,m} + \vec{\tau}_{\text{pass},j,m} . \tag{13}$$

This gives us a set of $J$ linear equations, where $J$ is the number of joints

$$\sum_{m=1}^{M} act_m \cdot \vec{\tau}_{\text{act},j,m} = \vec{\tau}_j - \sum_{m=1}^{M} \vec{\tau}_{\text{pass},j,m} , \tag{14}$$

for $j$ from 1 to $J$, which can be cast into the form of a matrix equation

$$T \cdot \text{act} = \beta, \tag{15}$$

where $T$ is a $3J \times M$-matrix of the torques $\vec{\tau}_{act,j,m}$ for all joints $j = 1, \dots, J$ and muscles $m = 1, \dots, M$,

$$T = \begin{pmatrix} \vec{\tau}_{\text{act},1,1} & \cdots & \vec{\tau}_{\text{act},1,M} \\ \vdots & \ddots & \vdots \\ \vec{\tau}_{\text{act},J,1} & \cdots & \vec{\tau}_{\text{act},J,M} \end{pmatrix}, \tag{16}$$

where $act$ is an $M$-vector of the activation of all muscles $m = 1, \dots, M$ ,

$$act = \begin{pmatrix} act_1 \\ \vdots \\ act_M \end{pmatrix} \tag{17}$$

and where $\beta$ is a $3J$-vector given by

$$\beta = \begin{pmatrix} \vec{\tau}_1 - \sum_{m=1}^{M} \vec{\tau}_{\text{pass},1,m} \\ \vdots \\ \vec{\tau}_J - \sum_{m=1}^{M} \vec{\tau}_{\text{pass},J,m} \end{pmatrix} . \tag{18}$$

Let us denote the set of solutions for the activation vector $act$ in the matrix equation (15) by

$$\mathcal{A} = \{ act \in [0,1]^M | T \cdot act = \beta \}, \tag{19}$$

for given $T$ and $\beta$. We can now try to minimize the overall muscle activation using MATLAB's linear least-squares solver `lsqlin`, (15), so that the optimized activation reads

$$\widehat{act} = \min_{act \in \mathcal{A}} \|act\|^2 \qquad (20)$$

with the usual Euclidean vector norm $\|act\| = \sqrt{act_1^2 + \cdots + act_M^2}$.
Instead of minimizing the muscle activation, though, we can also try to minimize the force that the muscles actively produce on the joints. For a given joint *j,* that force reads

$$\vec{F}_j = \sum_{m=1}^M act_m \cdot \vec{F}_{act,m} \cdot 1_{\mathcal{M}j}(m) \qquad (21)$$

with $\vec{F}_{act,m}$ given by (2.1). Hence, we optimize according to

$$\widehat{act}_{force} = \min_{act \in \mathcal{A}} \|C_F \cdot act\|^2, \qquad (22)$$

where $C_F$ is a $3J \times M$-matrix defined by

$$C_F = F/\|F\|, \qquad (23)$$

with the components of $3J \times M$-matrix $F$ being the spatial components of the forces $\vec{F}_{act,m} \cdot 1_{\mathcal{M}_j}(m)$ that are actively produced by the muscle $m$ on joint $j$,

$$F = \begin{pmatrix} \vec{F}_{act,1} \cdot 1_{\mathcal{M}_1}(1) & \cdots & \vec{F}_{act,M} \cdot 1_{\mathcal{M}_1}(M) \\ \vdots & \ddots & \vdots \\ \vec{F}_{act,1} \cdot 1_{\mathcal{M}_J}(1) & \cdots & \vec{F}_{act,M} \cdot 1_{\mathcal{M}_J}(M) \end{pmatrix}, \qquad (24)$$

and with $\|F\|$ indicating the usual matrix 2-norm (maximum singular value, also called the spectral norm) of $F$ (more explicitly, according to the general definition, $\|F\| = \sup_{x \neq 0} \|Fx\|/\|x\|$, where the supremum is taken over all vectors $x$ that are not the null vector). The reason for the renormalization was that the optimization algorithm works best when the inputs are of moderate scale and not orders of magnitudes different in scale. The specific choice of the norm is not relevant, here we chose the matrix 2-norm in analogy to the usual Euclidean vector norm. The reason for the renormalization was that the optimization algorithm works best when the inputs are of moderate scale and not orders of magnitudes different in scale. The specific choice of the norm is not relevant, here we chose the matrix 2-norm in analogy to the usual Euclidean vector norm. Then, we created a combined optimization matrix, $C_s$ where we gradually shifted between muscle activation optimization and joint force optimization, using optimization parameter $s \in [0,1]$:

$$C_s = \begin{pmatrix} (1-s) \cdot \mathbb{1} \\ s \cdot C_F \end{pmatrix}, \qquad (25)$$

such that we optimized according to

$$\widehat{act}_s = \min_{act \in \mathcal{A}} \|C_s \cdot act\|^2, \qquad (26)$$

where $s = 0$ results in the first criterion (minimizing muscle activation), and $s = 1$ results in the second criterion (minimizing joint forces). We varied the optimization parameter in 11 steps ($s = 0, 0.1, 0.2, \ldots, 1$) for each of the 2 movements (walking and squatting), and for each of the 6 subjects, resulting in 11·2·6=132 simulations.

## 2.3　Analyze the accuracy of predicted forces on the knee joint

As the body weight (BW) has a systematic effect on the measured and predicted knee forces, all forces were normalized to the body weight, i.e., force in BW = (force in N) / (body weight in N). For each simulation, we calculated the root mean squared error (RMSE) as well as the mean error between the predicted and the measured knee contact-forces across the movement. To compare the measured and predicted joint forces, we calculated the mean trajectories with standard deviations across subjects.

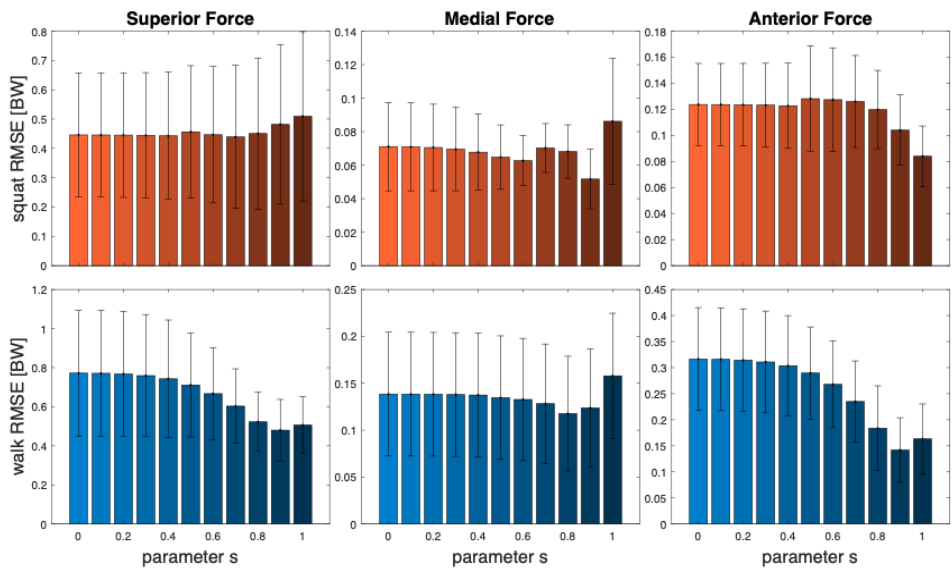

Figure 2: Root mean squared error (RMSE) between the measured and the predicted joint forces during squatting (top) and walking (bottom), averaged over the participants, for different parameters $s$. A parameter $s = 0$ minimizes the muscle forces, while $s=1$ minimizes the joint forces. The following knee joint forces are shown: superior force (left), medial force (middle), and anterior force (right). Error bars indicate one standard deviation. For $s \approx 0.9$, i.e., a minimization of predominantly joint forces, the RMSE is minimal.

## 3.　Results

The Myonardo-model was able to predict the measured knee forces with an accuracy of <0.5 BW in the superior direction and about 0.1 BW in the medial and anterior directions. The prediction accuracy of the measured knee joint forces depended on the optimization criterion $s$, with errors ranging from 0.03 BW to 1.1 BW (Fig. 2). The minimum RMSE values between the measured and the predicted knee forces during walking and squatting were found for $s \approx 0.9$. For this value of $s$, the mean RMSE between the predicted and the measured knee forces for squatting movements were: 0.44 BW superior, 0.05 BW medial, and 0.08 BW anterior. During level walking, the RMSEs were: 0.48 BW superior, 0.12 BW medial, and 0.14 BW anterior.

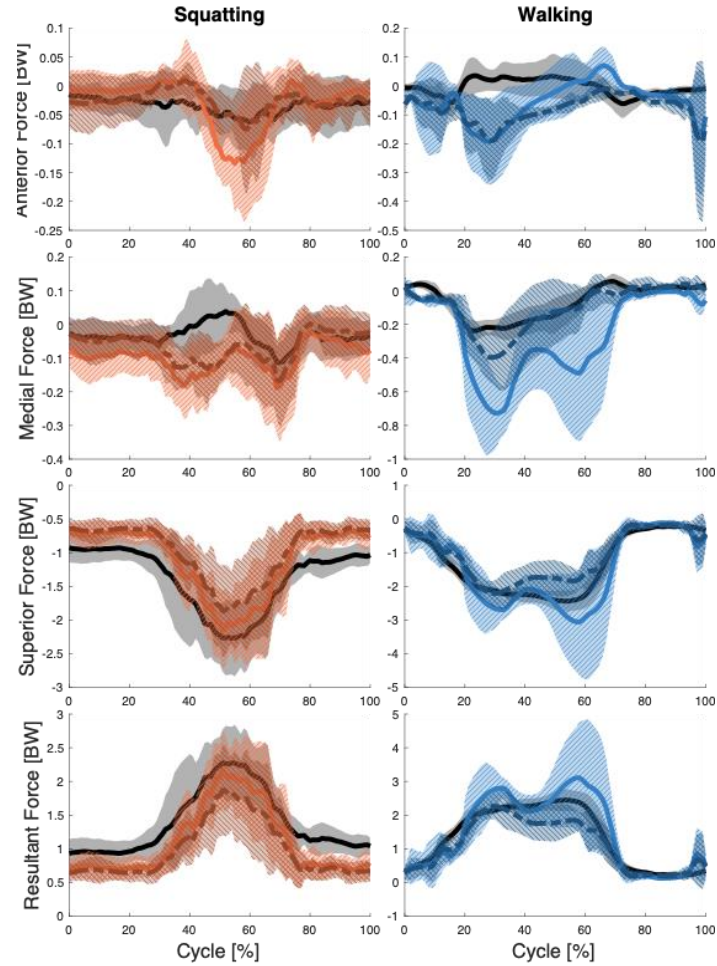

Figure 3: Comparison of the measured knee joint forces (black with gray shading) with the predicted knee joint forces (colored with hatched shading). Predicted forces are shown for optimization parameter s = 0 (bright-colored solid curves, upward hatching) and $s = 0.9$ (dark colored dashed curves, downward hatching) during squatting (left, red) and level walking (right, blue). The lines show the mean values across all subjects and the shaded areas show the corresponding standard deviations.

The predicted resulting forces were generally underestimated. Minimizing for muscle activation (s=0), the mean difference between the predicted and the measured knee forces for squatting is $-0.28 \pm 0.32 BW$ and for walking $0.15 \pm 0.27 BW$, and even more pronounced while minimizing for joint forces (s=1) for squatting with $-0.45 \pm 0.33 BW$ and for walking $-0.26 \pm 0.22 BW$. Nevertheless, the standard deviations of the predicted and measured trajectories were mostly overlapping (Fig. 3). Compared to the superior forces, the medial and anterior forces were much smaller. The largest deviations between predicted and measured medial forces were found during the first half of the gait cycle.

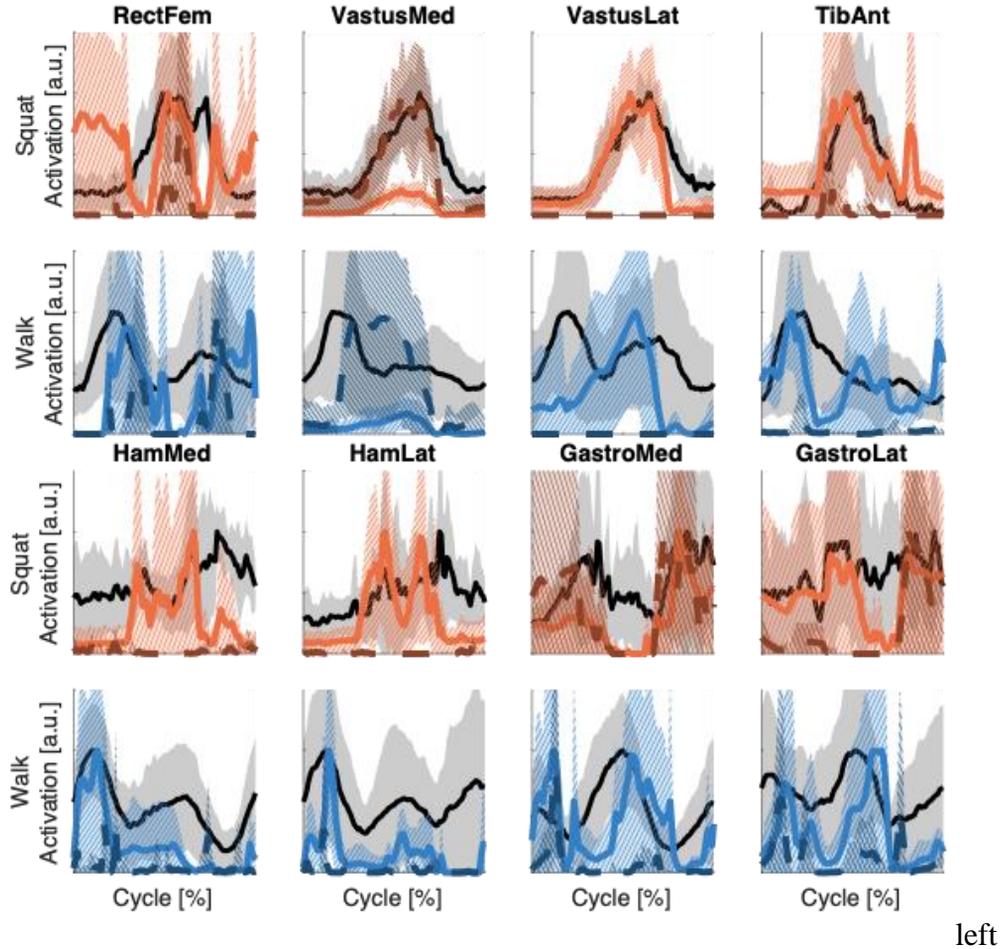

left

Figure 4: Measured and predicted muscle activation for squatting (first and third row) and walking (second and fourth row). The measured EMGs (black, solid grey hatching) were averaged over the six patients, and normalized to the maximum value over all values of s for each muscle separately. The predictions while minimizing the muscle force (s = 0, bright solid line, upward hatching), and minimizing the joint force (s = 1.0, dark dashed line, downward hatching) are shown.

How do different optimization criteria influence knee joint forces? The optimization criteria that were used to calculate the predictions have a large influence on the resultant knee forces $F_R$. Minimizing for muscle activation (s=0) results in maximum knee forces of $F_R = 4.03 \pm 1.25\ BW$ for walking, and $F_R = 2.49 \pm 0.48\ BW$ for squatting, while minimizing for joint forces (s=1) results in maximum knee forces of $F_R = 2.25 \pm 0.38\ BW$ for walking, and $F_R = 2.12 \pm 0.65\ BW$ for squatting. So, changing the optimization parameter leads to a reduction of -44.2% in walking (-1.78 BW), and a less pronounced reduction of -14.8% during squatting (-0.37 BW). This shows that it is possible to optimize the control of leg muscles to reduce joint forces, even though net torques for the given movements are still produced (Fig. 3).

The optimization to reduce the joint forces during walking and squatting results in a deactivation of bi-articular muscles (Fig. 4: e.g., rectus femoris, gastrocnemius). This reduction of force is partly compensated by uni-articular muscles (Fig. 4: e.g. vastus medialis).

## 4. Discussion₃Error! Reference source not found.

The aim of the current study was (1) to assess the effect of different movement strategies in a computational musculoskeletal model on the predicted forces in the knee joint during squatting and walking, and (2) to assess the validity of these predicted forces by comparing them with real-life data from instrumented knee prostheses. We changed the optimization criterion stepwise from a focus on minimizing muscle activation to minimizing joint forces and found that both the prediction error and the predicted joint forces depended on the optimization criterion that was used. This indicates that an adaptation of movement strategies can reduce knee joint forces, which could aid, e.g., people who experience load-related joint pain.

The model predicted the measured forces with an average RMSE of about 0.22 BW. The accuracy of the predictions of the measured knee joint forces depended on the optimization criterion that was used. We found the highest RMSE in the superior direction, both during squatting and level walking. This is likely related to the relatively high superior forces during these movements (about 2-2.5 BW, as compared to ~0.2 BW and ~0.1 BW in anterior and medial directions respectively).

We found that muscle coordination highly influences knee joint forces, as the choice of optimization parameter led to a reduction of up to -44% in walking and -15% in squatting, relative to the max predicted joint force. Patients with injuries such as damaged articular cartilage, osteoarthritis, osteoporosis, or joint inflammation, perceive a pain that is correlated with joint loading. These patients are likely to adapt their movement strategy to reduce the joint forces during locomotion or other daily life activities.

The CAMS-knee data set also reports surface EMG for eight muscles in both legs (Taylor et al., 2017). Due to electrical resistance at the skin and subcutaneous fat, muscle movements with respect to the electrodes and the distribution of distances of single motor units to the electrodes, EMG signals represent just an indirect and noisy measure of muscle activation. In the present study, we therefore compared the predicted muscle activations and the measured EMG signals only qualitatively (Nejad et al., 2020).

We found that joint force optimization led to a reduced recruitment of bi-articular muscles. This can be explained as bi-articular leg-muscles generate a flexion torque in one joint and an extension torque in the other, or vice versa, e.g., an activation of the bi-articular biceps femoris muscle generates an extension torque at the hip joint but also a flexion torque at the knee joint, that does not contribute to the necessary overall leg extension. These counteracting torques must then be compensated by extensor muscles at this joint, leading to increased joint forces. Therefore, to minimize joint forces, the optimization algorithm will result in a muscular control pattern that favors uni-articular over bi-articular muscles, to keep co-activation low (Giddings et al., 2000). Over time, these changes in muscle activation patterns may lead to muscular atrophy of bi-articular muscles and a hypertrophy or chronic overload of some uni-articular muscles.

To our knowledge there are just a few studies that compare in vivo measured knee forces with the predictions of musculoskeletal models. The in vivo data in instrumented knee implants were measured by Fregly et al. (2012), Kinney et al. (2013) and Taylor et al. (2017). Marra et al. (2015) and Zhenxiana et al. (2016) developed a subject-specific full lower limb Anybody model with an RMSE between 25% and 36% BW. The same data sets were compared to different OpenSim models (Curreli et al., 2021) and different optimization techniques (Knarr & Higginson, 2015) with RMSE ranging from 36% BW and 62% BW. Two studies compared their predictions based on generic OpenSim models with the data of Taylor et al. (2017) that were also used in the present study. Schellenberg et al. (2018) and Imani Nejad et al. (2020) predicted the knee joint forces during squatting with a RMSE of 58.5% BW and 105% BW, respectively, and a RMSE of 47% BW during walking (Imani Nejad et al., 2020).

Compared to these predictions, the results of the present study are comparable or better, which is probably due to the optimization criterium to minimize joint forces. The errors in our model can be decreased even further by slightly increasing muscular co-contraction, thus increasing the currently underestimated joint forces (cf. figs. 3, 4). However, since in this study, our aim was not to produce the best possible fit to the measured joint forces, but rather to analyze the influence of the optimization on the error, we did not present these results. Such an increased co-activation is likely to occur in practice, e.g., to increase joint stiffness and self-stability (Giesl et al., 2004; Wagner and Blickhan, 2003), and would increase overall joint loading.

Furthermore, analyzing the optimized muscle activation pattern to reduce joint forces during walking and squatting, our data show that a deactivation of bi-articular joint muscles is related to compensatory activation of uni-articular muscles of the knee joint. Both optimization criteria used in the present study, i.e., minimizing muscle activation and minimizing resulting joint load, might not be the only cost function that is minimized in persons with injuries or pain in real-life situations. Interestingly, the optimization parameter *s* leads to a minimum RMSE between the predicted and the measured knee forces, for values of $s > 0.6$, indicating that the participants in the CAMS study used a muscular co-ordination pattern that minimized joint force. As these participants had a history of load-depending pain, they might have already shifted to this strategy as a way of pain avoidance. Unfortunately, this assumption cannot be proven at present, as it is still not possible to directly measure these joint forces in a healthy subject during everyday movements.

We simulated "movement strategies" as computational optimization criterions to obtain numerical solutions for our vastly redundant neuromuscular model. It is quite clear, that the nervous system does not perform this kind of optimization computations explicitly. Rather, it makes use of nervous control strategies such as desired length and contraction velocity of muscles (e.g., Latash et al., 2010, de Lussanet et al., 2012) or the avoidance of joint pain. The latter can be approximated by the criterion of minimal joint force used in the present work.

This study was carried out as a non-controlled pilot study to compare model-based prediction using two optimization criteria with real-life data from six patients after knee arthroplasty. Both optimization criteria (minimizing the muscle activation and minimizing the resulting joint load) might not be the main cost function that is minimized in persons with injuries or pain in real-life situations. For such populations, afferent proprioceptive information could be the basis for the optimization.

We found differences in RMSE between the three different force directions for the knee joint force, which may indicate that optimization is direction-dependent, or there are missing parameters when RMSE is higher, although it is not clear whether a higher RMSE, suggests lower accuracy to the calculated forces, is random or caused by an incomplete musculoskeletal model.

## 5.   Conclusion

Physical therapy interventions for people with movement related pain should consider specific exercises for bi-articular muscles, and computational musculoskeletal models may be used as a tool to analyze joint forces during locomotion and typical everyday movements. Patients may adapt their neuromuscular activation pattern to reduce joint forces during locomotion or everyday movements.

The direct comparison of forces measured in an instrumented knee prosthesis and those calculated using a full body musculoskeletal model are in good agreement with respect to all three components of the local coordinate system for both squatting and walking. The simulations confirm that the muscular contractions have a major contribution to joint load. They also showed that the choice of optimization criterion (minimal muscular force vs. minimal joint load) has a considerable influence on both prediction error and joint load. The computed joint forces were generally lower than the measured ones, which may be explained by a certain base level of co-contraction in the measured patients.

# 6.   Disclosures

# 7.   Acknowledgments

We thank Predimo GmbH for providing the Myonardo software. We thank Prof. Dr. William Castro for fruitful discussions. We are grateful to the CAMS-Knee Database (represented by the Laboratory for Movement Biomechanics of ETH Zürich, and the Julius Wolff Institute of Charité -- Universitätsmedizin Berlin) for kindly providing the data.